# "For an App Supposed to Make Its Users Feel Better, It Sure is a Joke" - An Analysis of User Reviews of Mobile Mental Health Applications

MD ROMAEL HAQUE and SABIRAT RUBYA, Marquette University, USA

Mobile mental health applications are seen as a promising way to fulfill the growing need for mental health care. Although there are more than ten thousand mental health apps available on app marketplaces, such as Google Play and Apple App Store, many of them are not evidence-based, or have been minimally evaluated or regulated. The real-life experience and concerns of the app users are largely unknown. To address this knowledge gap, we analyzed 2159 user reviews from 117 Android apps and 2764 user reviews from 76 iOS apps. Our findings include the critiques around inconsistent moderation standards and lack of transparency. App-embedded social features and chatbots were criticized for providing little support during crises. We provide research and design implications for future mental health app developers, discuss the necessity of developing a comprehensive and centralized app development guideline, and the opportunities of incorporating existing AI technology in mental health chatbots.

## 1 INTRODUCTION

One out of every four persons on the world has been impacted by mental or neurological issues at some point in their lives [82]. Mental health issues affect around 450 million people globally, making them one of the primary causes of ill-health and disability according to WHO [82]. In 2019, 20.6% of adults in the United States (51.5 million individuals) were affected by mental illness, representing 1 in every 5 adults and a sharp increase of 4% in just three years [57]. Due to the inaccessibility and high cost of traditional treatment, around 55% of people with severe mental illnesses do not receive treatment [91]. With the advancement of mobile technologies in the last ten years developers recognized a strong promise for digital tools like mobile phone applications to expand better accessibility at low cost to mental health (MH) treatment and services [9]. Prior study has acknowledged this breakthrough for making MH treatment more accessible, convenient, and adaptable to the patient's lifestyle [19]. By 2018, there were over 10,000 MH and wellness apps available for immediate download [114], with services ranging from symptom tracking and monitoring to implementing scientifically grounded therapy, such as CBT and Mindfulness, as well

Authors' address: MD Romael Haque, mdromael.haque@marquette.edu; Sabirat Rubya, sabirat.rubya@marquette.edu, Marquette University, Milwaukee, WI, USA, 53233.

421

as various interactive tools and modules for self-guidance [8]. However, prior study has shown that while access and availability of MH applications has increased, concerns about privacy, effectiveness, and usability have also been raised [113]. According to a recent study, 7–42% of users continued to use the apps after four weeks, but just 0.5–28.6% after six weeks [87]. The use and adherence with MH applications in real-world settings may range between 1% and 29% [39]. Furthermore, because the majority of the applications accessible are not evidence-based, their efficacy in aiding people with MH concerns is questionable [8].

Prior research has attempted to understand these challenges using data from a variety of sources, including the descriptions of applications in mobile app stores [29], publicly available peer-review databases [8], frameworks and evaluation criteria from engineering and informatics literature, and interdisciplinary organizations [20], etc. Smartphone-based apps represent a unique opportunity to expand the availability and quality of MH treatment, and the COVID-19 pandemic caused a surge in MH and wellness app downloads [119]. But only a few studies focused on user-centered design and usability evaluation of these apps. There is a gap in HCI and CSCW literature in understanding the limitations of these apps through a user-centered lens. To gain an empirical understanding of why there is such a low level of engagement and significant dropout rates, it is vital to examine the challenges and concerns regarding these apps from the consumers' perspective and their real-life experience.

User reviews provide a unique venue to extract information around usability issues and evaluate efficacy [76]. Studies have demonstrated that analyzing user evaluations can provide in-depth knowledge of user experiences. It can assist in identifying potentially harmful functionalities, incorporating community values and expectations into app design in a variety of domains, including mobile health technologies [2]. Furthermore, users consider these reviews as a credible source to learn about expectations and potential consequences, allowing them to make informed judgments about what to choose and avoid while thinking about these apps as treatment supplements. In this study, we utilize the app descriptions and user reviews of the mobile MH apps from two most popular app stores to investigate two research questions:

**RQ1:** What is the state of the art in mobile MH applications, in terms of targeted MH problems and techniques used to support users?

**RQ2:** What are the concerns and perspectives expressed in user reviews published on mobile app store platforms about the usability and efficiency of mobile MH applications?

To answer these questions, we used inductive thematic analysis to examine user reviews of 164 apps from Google Play Store and Apple App Store. This publicly available data (user reviews) provides in-depth analyses of users' personal app experiences. We collected user reviews of mobile applications from online that offer support and treatment for a variety of MH concerns, including (but not limited to) depression, anxiety, stress, and other MH concerns. We found that there is a lack of attention in these apps to offer in real-time and emergency support, as well as a lack of transparency in collecting private data from users. There were persistent complaints about the app-embedded online communities regarding inconsistent moderation approaches and biases against specific genders, races, and other marginalized groups. The app-embedded chatbots often failed to deliver helpful and practical solutions due to the inability to generate personalized responses.

CSCW has a long history of research around the nature of analyzing the efficacy of technologies related to the treatment of mental health care [38, 93, 97], health ICT policies and agendas [63, 85], and designing collaborative support system for the users [73, 79, 96]. Recent studies in CSCW [38, 88, 103] pointed out the necessity of assessing socio-technical perspectives of such technologies from users' viewpoints. These studies have emphasized the importance of user-centric design and evaluation of systems targeted for vulnerable populations such as individuals suffering from mental health concerns. [72, 107, 117]. We contribute to this body of work by providing concrete research and design implications for future MH app developers by considering the limitations of existing apps found from the user-feedback. Our findings point to the need for a comprehensive guideline for MH app developers. User reviews are one of the first sources people seek for when installing the apps [59]. Developing guidelines with user reviews in mind may help users make informed decisions, and bridge the communication gap that exists between developers and users.

## 2 RELATED WORK

We situate our work in previous literature by discussing why and how researchers conducted user review analysis, introducing the current state of the art research on MH apps, and describing the necessity of following guidelines in developing MH apps.

### 2.1 User Review Analysis

Online reviews are a type of product information generated by individuals based on their own personal experiences with the product. Consumers can post product reviews with content in the form of numerical star ratings and open-ended customer-authored remarks about the product on online sites [71]. User reviews are an integral component of the user experience, especially when topic satisfaction is crucial for adoption [59]. Like many other online marketplaces, mobile app stores allow users to post their feedback based on their experience of using the apps [110]. Users may utilize this feedback loop to evangelize and promote apps they like, as well as alert other users of the apps' shortcomings. Given the highly competitive nature of the mobile app market, customers consider user reviews an important resource to take informed decision about which apps (and what features in those) will or won't work for them [76]. In recent years, researchers have offered a wide range of data mining tools to help developers and analysts use this feedback [46]. In terms of analytical tools and methodologies to extract the rich contextual information, they used sentiment analysis [10], SVM [121], Naive Bayes [84], maximum entropy classification [52, 121], N-gram model [121], and qualitative analysis methodologies [110] etc. These methodologies were utilized to identify relationship between ranks and sales [21, 23], explore the underlying factors of customer value [83], compare reviews of different products in a category to find the reputation of the target product [49], create summary of opinions for question answering [18], identify fake reviews [66], exploring user characteristics [40, 42, 60] etc. from different e-commerce platforms such as amazon [10, 84], mobile app stores [3, 105], hospitality sectors [33], restaurant sectors [61] etc. There are few works that employ user evaluations to investigate mobile mental health/well-being/behavior change applications for future design opportunities [28], figuring out usability concerns or user characteristics [2].

Our research is motivated by this body of work that considers user-generated feedback as a valuable source of information [90] and make useful meaning out of them [76]. User-generated content allows us to gather information from a wide spectrum of people that is difficult to collect with typical data collection methods [90, 109]. Through this investigation we will develop a deeper understanding of the real-world experience and usability concerns of mental health app users and extract the complaints and suggestions including problem reports, feature requests, etc. All these are valuable resources for app developers to improve user experience and satisfaction [83].

### 2.2 Mobile MH Technologies

The extensive adoption of mobile health (mHealth) technologies is unprecedented in the history of healthcare and access to treatment [47]. Apps particularly addressing mental health (MH) issues [47] target a broad range of psychological concerns including anxiety [106], depression [101], schizophrenia [100] etc., and vary in design and functionality. They offer a variety of functions, including assessing, tracking and monitoring of illness symptoms [36], providing resources and education to mitigate and manage symptoms that arise as a result of a mental health issue, such as breathing [22], meditation [45], etc., game–based apps that help to manage anxiety and achieve healthy living [14], and facilitating support through a community of peers and/or coaches [41]. Despite the fact that mental health apps have been shown to be beneficial in supporting mental health, there is still a significant number of issues and concerns, such as early dropouts [68], low engagement [74], and not being very effective in improving the condition of the patients [13]. For example, researchers found clinically proven methods such as CBT and positive psychology often do not perform as expected in mobile-based platforms [75]. Moreover, prior research showed different self-guided mobile applications suffer from early dropouts [6]. Many mental health applications report low engagement and a high attrition rate. However, there is little evidence to support these claims [112]. Possible reasons include usability issues, such as software bugs, unappealing user interfaces, data loss, battery and memory usage issues, lack of guidance and explanation, etc. [3]. Additionally, due to privacy concerns, people are often wary of using mental health apps [108]. Very few evidence and research based applications that are available often fail in implementing important features, such as understanding early warning signs and triggers [69], wellness plans [113], etc. These result in a lack of trust among users and healthcare providers, as well as poor engagement and dropouts [68, 74].

The importance of applying best practices such as making comprehensive support accessible and available inside the app, as well as following clinical evidence-based approaches has been acknowledged in previous research. However, research into the effectiveness of mental health apps is still in its infancy, and there are ongoing discussions about how best to examine the efficacy of mental health apps. Hundreds of apps are being developed and are downloaded from the app stores without significant research to confirm the effectiveness of such apps. A comprehensive understanding of user perspectives and experiences may aid in answering research questions around this debate and inform design of effective apps. A qualitative analysis of app store reviews will contribute to an understanding of the current confusing, uncertain picture about consumer interest in using MH apps in the real world (as opposed to research settings).

### 2.3 Guidelines for Developing Mobile MH Applications

Anyone who wants to create or recommend an MH app will confront a number of challenges in terms of comprehending the design elements and functionalities [62, 70]. Even experts in this field, such as healthcare professionals, have difficulty determining what will and will not work for their patients [8, 27]. Even if developers and professionals try to anticipate all of these technologies' possible ramifications and create the recovery process in such a way that they may be avoided, it doesn't always work out in the best interests of the users [27, 51]. Prior research showed that users risk serious consequences as a result of the lack of evidence-based development of such applications, as well as poor design and functionality implementation, such as their mental health concerns becoming even more triggered [32]. Recognizing these issues, different agencies offered directions in establishing guidelines for developing these applications, such as government bodies that regulate health apps have begun to issue guidance to explain the law, and offers suggestions on how to create apps that run in accordance with ethical values [118]. Self-regulatory ethical codes have also been produced by professional organizations, informing and instructing app developers on industry norms for professional behaviour [5, 7, 30]. The Federal Trade Commission (FTC) has banned the promotion of health apps that made unfounded claims regarding their intended use [24, 25]. They proposed an interactive tool to assist app developers with the knowledge of legal compliance [37]. The purpose is to provide developers a quick overview of a few key federal laws and regulations from different agencies [37]. Besides, different approaches to establishing guidelines have been noted, such as the American Psychiatric Association (APA) proposing a learning tool called "App Advisor" to help psychiatrists and other mental health professionals navigate appropriate applications for their patients by highlighting important aspects they should consider [4]. Another non-profit, "Psyberguide," works to expand access to mental health resources for those who didn't have it before, as well as to help users learn how to utilize technology to better their well-being [81]. They assessed a number of mental health apps using three criteria: credibility, user experience, and transparency [81]. Both the APA and Psyberguide guidelines are based on the opinions of healthcare professionals and experts [4, 81].

However, there exist several shortcomings in the existing guidelines for the design and development of MH apps. First, many of these focus only on ethical norms and notify developers about legal compliance. The rules aren't very detailed or user-centric. There is no way to make sure if any of the the components is not misunderstood by the developers [89]. Furthermore, only adhering to ethical standards and legal requirements does not guarantee that the app will not cause any negative repercussions for the users [15]. Second, each of these guidelines follow a different set of criteria and do not provide a standardized set of recommendations for app developers. App developers may not be aware of all available legal and professional recommendations and policies, nor they will have access to these. This is due in part to the fact that relevant advise is handled by a variety of organizations, the majority of which are independent, across a wide range of industries [86, 92]. Since the guidelines are dispersed in so many areas, and are subject to change with new apps being introduced, it is difficult for the app industry/ developers to keep track of the latest set of guidelines [111]. Third, only a few, such as "Psyberguide," offer guidance to users in deciding between different applications, but these do not fully consider actual user feedback and real-world experiences, but rather on the opinions of professionals and experts in the most part. Integrating users' opinions in critical in offering app suggestions in such a sensitive health context [110].

These shortcomings call for a standardized all-in-one guideline for MH app development, containing effective techniques to avoid potentially harmful functionalities and to include community values and expectations while designing the apps. It should provide the patients with knowledge of expectations and possible ramifications, and enable them to make informed decisions about selecting apps. This work aims to build a comprehensive understanding of the state of the art in mobile mental health-related apps and their user feedback from the two largest online app store marketplaces, that may inform the design of such a standard app development guideline.

## 3 METHODS

Users of mobile apps can download, review, and comment on the apps depending on their experience and satisfaction. This publicly accessible data (user reviews) offers in-depth evaluations that include positive, negative, and neutral feedback. Written evaluations can also reveal customer satisfaction with certain program features or desire for enhancements. As a result, reviews are regarded as a significant source of information.

### 3.1 Selection of Sample Apps and Reviews

We searched for potential apps focused on MH on two dominant mobile platform app stores (Google Play for Android and Apple App Store for iOS). We used the search term "Mental health apps" on both app stores. The search was conducted from the homepages of the app stores without logging in to any specific account. This step was taken to ensure that no ranking algorithm could be applied by the system to prioritize any specific user preference. Therefore, although the search results may not be fully comprehensive (as found from convenience sampling), these apps represent the sample in (almost) the same order that consumers would likely be exposed to and thus, most likely to use. The initial inclusion criteria were: an app must have an English-language interface and must have at least 100 downloads. These two filters gave us a list of 249 apps from Google Play and 117 apps from Apple App store. The authors carefully read the app descriptions and observed the screenshots of the app features to make sure if an app targets at least one mental health condition. Apps were excluded if they fell into any of the following categories: does not target a mental health issue, has less than 100 reviews (very few reviews would not give us enough useful or representative opinions), or is not in English. The final list included a total of 117 applications from Google play store and 76 applications from apple app store. Twenty nine applications were compatible with both platforms. As a result, a total of 164 applications were selected. Among the 164 applications analyzed in our study, 68 applications can be downloaded for free, 93 applications can be downloaded for free but an in-app purchase is required to access all functionalities, and 3 applications are being offered as a paid product upfront. The "Supplementary Material" contains a complete list of apps considered in our study.

We wrote python scripts to scrape application details and all available user reviews from the 189 apps. In order to extract recent critical user feedback of the apps we applied the three following inclusion criteria for filtering.

- Timeline: We included reviews that were published between January 1, 2019 and May 1, 2021. As the marketplaces change rapidly with new apps and updates of the existing apps, we wanted to capture the most recent user reviews. In addition, this timeline would allow us to understand if there exists any substantial difference in the reviews between pre and during COVID-19 period.
- Length: We included reviews with character counts of 200 or more. According to a previous study, the average amount of characters in a relevant review is 110.8. [66]. We chose 200 characters for the scope of our study because reviews with more characters will aid us in uncovering deeper insights when performing the analysis. Moreover, shorter reviews often tend to be fake and are created for promotional purposes. Ignoring those reviews assisted us in improving data integrity.
- Star rating: This filter was used to rank the reviews in an order that would allow us to arguably differentiate the expressed critical or negative insights from the positive ones. We categorized the reviews in two classes: reviews that have a one, two, or three star rating and reviews that have a four or five star rating. For discovering the negative insights discussed in the reviews, we considered all reviews with <=3 stars and reviews with >3 stars that contain some negative components in them. We took into consideration the positive components in the four star and five star reviews to understand users' positive opinions or preferences about different aspects of the apps.

2159 reviews from the Google Play Store and 2764 reviews from the Apple App Store met all of the inclusion criteria. These reviews are based on 117 apps from the Google Play Store and 76 from the Apple App Store. All of the reviews have a unique coding system that can be easily traced back to the application and platform from whence they emerged. During the analysis, the lead author was responsible for carefully reading each review and ensuring that all personally identifying information was replaced or removed.

### 3.2 Data Analysis

First, to gain an understanding of the current status of the commercially available MH apps, we categorized the 164 unique applications based on what mental health condition(s) they target and what intervention tools and techniques they adopt. The findings from this categorization are described in section 4.1. Then, we analyzed the reviews (filtered with the inclusion criteria) using an inductive thematic approach [77]. Thematic analysis was chosen because it allows systematic study of a big data collection and aids in the understanding of patterns in the text while taking context into consideration [56, 110]. Previous research [110] has shown that data obtained from thematic analyses may be examined and reported qualitatively, therefore we employed a qualitative thematic analysis. Analysis consisted of two passes. In the first pass we generated open codes (over 40 total) to capture diverse viewpoints from the reviews. This high number of open codes is due to our effort to capture the nuance in the specific insights included in each review and was significantly reduced through the process of memoing and clustering [77]. In the second phase of analysis, we memoed and clustered the codes using a constant comparison method operationalized as affinity mapping. Each open code was compared to others and positioned to reflect its affinity to emerging themes and clusters. The reported themes consisted both of ones that appeared consistently across multiple reviews, and also the ones that came from reviews that represented divergent responses and opinions. Despite the fact that we first thematically analyzed the negative content of all the reviews, we did not limit ourselves to only looking at the critiques. In section 4.3, we provide a brief description of what aspects of these applications users liked and reviewed positively as found from our analysis of the positive components of the >3 star reviews.

### 3.3 Data Integrity and Ethical Considerations

App stores, like many other online have reviews posted by fake and paid users. However, prior research [66] showed that in "Health & Fitness" category the percentage of potential fake reviews is very low (about 6%). Fake reviews also tend to have good star ratings [66]. Since we are mostly considering reviews with three or fewer stars, we assume that almost all of the reviews included are original and can be trusted.

Apple website has an automatic scraper detector that blocked our script several times as it was requesting from the same IP address within a short period of time [48, 116]. We circumvent this by identifying those apps where the review collection was incomplete and running our scripts on those apps iteratively from different IP addresses. We ran an additional script to detect and eliminate all the duplicate reviews (if present).

It is unethical for researchers to use any personal information from the internet if the data or information is restricted to a certain group of people or communities [122]. Hence, we ensured the webpages we collected the data from are public and not restricted to certain communities or populations. [122] also adds to this conversation of ethical implications by emphasizing that more than expected information should not be revealed through the combination of visual and textual elements. To ensure that, we intentionally refrained ourselves from publishing or revealing any identifiable information that were shared in those pages even though they were public.

### 3.4 Positionality

The research team consisted of members whose family members and close relatives have suffered from MH issues. One team member is an expert in intersectional stigma, online social support, and MHcare, while another is an expert in working with marginalized and vulnerable populations.

### 3.5 Limitations

Our analysis methods and selection criteria have some inherent limitations. First, we may be missing feedback from users who are not comfortable (or do not care) about writing their experience on online platforms. However, since we considered a representative sample of the MH apps and the number of reviews were high, we can safely assume that the concerns we found will be representative of user opinions. Second, we elected not to incorporate critical assessments around broad UI issues and software defects in our user-centered evaluations, largely because we aimed to capture a broader picture of customers' real-life experience and concerns around the usability and efficiency that a mere software update wouldn't be able to solve, and need more attention from the researchers and the app developing organizations. Third, for this study, we looked at ratings from only the two major mobile platforms (Google and Apple). Other mobile platforms were not taken into account for the scope of this research.

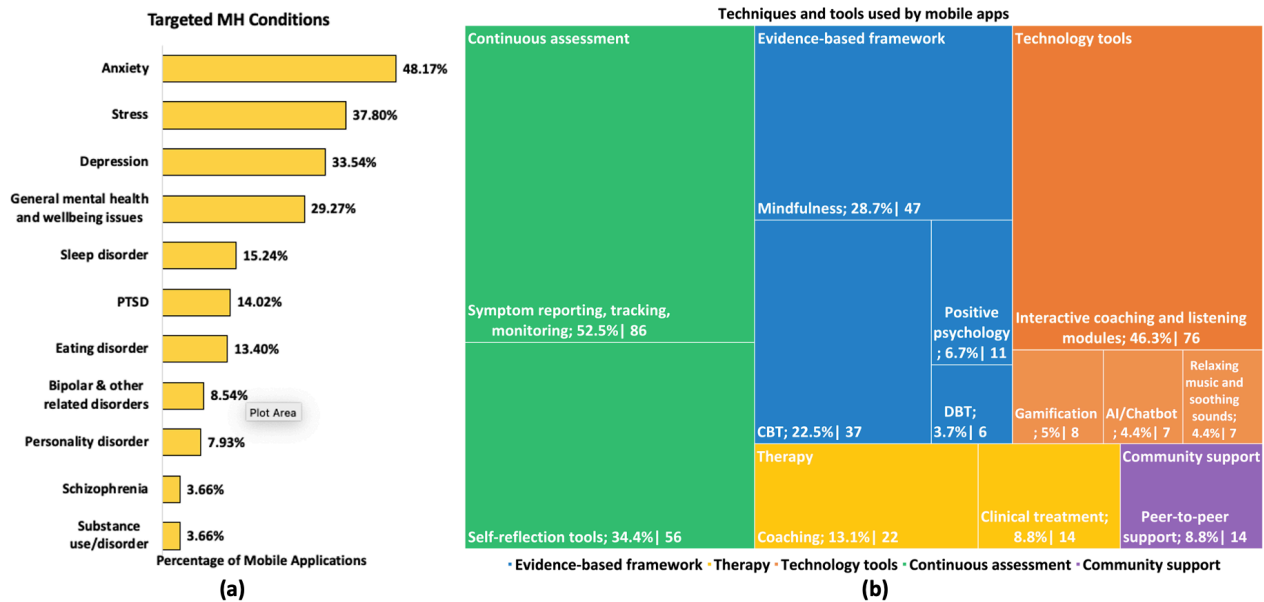

Fig. 1. **(a) Targeted MH conditions covered by the (percentage of) apps considered in our study. (b) Frequency distribution of the apps in terms of techniques and technological tools used.**

## 4 FINDINGS

We provided our findings from the analysis in this part divided into two subsections. In the first subsection, a descriptive analysis of the current state of the overall mobile mental health (MH) app market is presented. In the second subsection, perspectives and sentiments raised by users such as inconsistent moderation, toxic social interaction, lack of transparency and useful functionalities regarding chatbots are presented.

### 4.1 State of the Art in Mobile Mental Health (MH) Applications

We begin with an overview of the apps that were selected from our careful consideration following the approach described in section 3.1. Although the list of apps may not be exhaustive, it is a large representative sample of the existing mobile MH apps from two most popular app stores. Thus, we can assume that the descriptive statistics can be generalized and the findings reflect the current status of the overall mobile MH marketplace. We provide categorizations of the apps on two key aspects: types of MH concerns they target and techniques and technologies used to assist users.

*4.1.1 Targeted MH conditions.* 1(a) shows the number and percentage of apps per addressed MH problems, sorted in descending order. These categories are not mutually exclusive, which means that an app can address multiple MH conditions. Anxiety is the most commonly addressed issue (79, 48.17%). Other significant issues are : stress (62, 37.8%), depression (55, 33.54%), general mental health and wellbeing (48, 29.27%), sleep disorder (25, 15.24%), PTSD (23, 14.02%), eating disorder (22, 13.4%), bipolar & other related disorders (14, 8.54%), personality disorder (13, 7.93%), schizophrenia (6, 3.66%), substance use/disorder (6, 3.66%).

*4.1.2 Intervention techniques and approaches used.* Different apps use different strategies and tools to treat various MH issues. These techniques, as found from the description of the applications covered in our study (n=164), can be split into five broad categories: evidence-based framework, therapy, technology tools, continuous assessment, and community and peer support. Figure 1(b) represents the number and percentage of these broad categories and subcategories. These categories are mutually inclusive, implying that an app can implement a variety of strategies.

- **Evidence-based framework:** A handful of the apps implemented techniques that are evidence based and scientifically grounded such as CBT, DBT, mindfulness and positive psychology. Assisting people in achieving mindfulness is a prominent evidence-based strategy. We found 47 (28.7%) apps utilizing mindfulness. The ability to be fully present, aware of where we are and what we're doing, and not unduly reactive or overwhelmed by what's going on around us is known as mindfulness. Meditation is a popular method for achieving mindfulness. These apps offer guided meditations and courses, sleep meditation, breathing exercises, stress relief and coping meditations, exercises related to finding focus building productivity, tailored reminders about meditation, and the option to measure the number of minutes meditated, etc. The duration of these courses and guides might range from a few minutes to several hours. CBT (cognitive behavioral therapy) is an another type of psychotherapeutic treatment that teaches people how to recognize and change harmful or distressing thought patterns that affect their behavior and emotions. Among the applications we considered, 37(22.5%) implemented CBT. Majority of the apps have a number of therapeutic elements such as tests and scales to assess one's well-being, resources about symptoms and treatments, pleasurable activities to improve one's mood, tasks to complete to practice new coping skills, communicating through inspirational quotes, and suicide prevention measures (providing links to support services, ability to prepare a crisis plan), etc. Positive psychology is a modified type of CBT that is essentially a cognitive treatment for changing negative thoughts and dysfunctional perceptions into more positive ones. Some of these apps (11, 6.7%) apps allow users to reflect on themselves using diaries and journals, as well as add affirmations to encourage positive thinking. Finally, DBT is another evidence-based technique that these programs use. DBT (Dialectical Behavior Therapy) is a modified version of CBT. Its primary objectives are to teach people how to live in the now, build healthy stress coping mechanisms, regulate their emotions, and improve their interpersonal connections. Only 6 (3.7%) applications used this technique.
- **Therapy:** This wide category was broken into two subgroups: coaching (22,13.1%), where users can connect with professional coaches and active listeners guide them through various mental and physical wellness recovery processes and clinical treatment (14,8.8%), in which users can remotely connect with a medical professional or a registered therapist and go through their therapy session with them. In few of these apps, users can be matched with coaches and therapists who suit their style and preferences using algorithms and matching systems depending on their needs. Remote therapy and coaching solutions are highly useful for patients who live in remote places or who have physical disabilities.
- **Technology tools:** We found four different types of technological tools and features in the apps we included. First, interactive coaching and listening modules (76, 46.3%) where app users interact with a dynamic tool based on their experienced symptoms, and receive suggestions on ways to improve their specific mental conditions. These resources offer suggestions for navigating around potential trigger symptoms, avoiding risky activities, and learning about different treatment options on their own. Second, gamification (8, 5%) is a technology intervention that has recently gained popularity. Various game-based apps assist users in assessing and coping up with symptoms through provision of virtual reward for setting and completing specific goals. Leveling up, collecting money, a map to display progress, rankings, and characters are just a few examples of the video game-inspired features. People's motivation is expected to enhance by rewards and incentives in particular. Third, AI/chatbot (7,4.4%), in which developers utilize artificial intelligence and machine learning techniques to create bots that can interact with users by talking to them, coaching, suggesting treatment, guiding them through exercises, and so on. Fourth, relaxing music and soothing sounds (7, 4.4%) that help people recuperate from various metal difficulties.

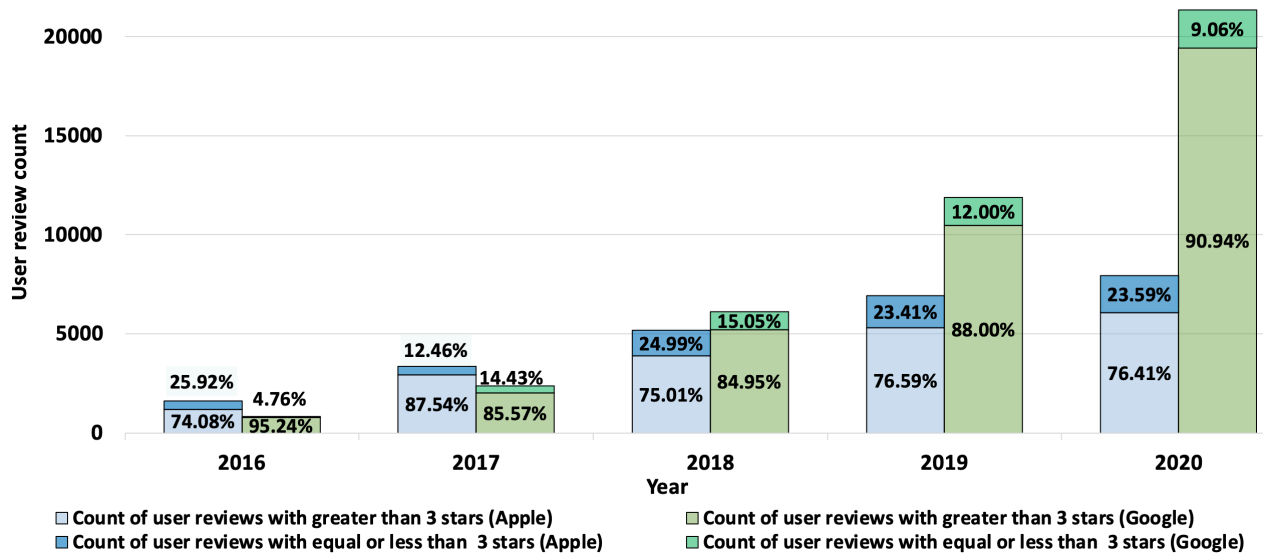

Fig. 2. **Number of user reviews published in Google Play Store and Apple App Stores in the last five years for the apps considered in this study**

- **Continuous assessment:** Continuous Assessment refers to approaches in which mobile apps assist users in better comprehending their present condition by contrasting recent symptoms with a history of previous symptoms. It requires continuous user interaction, as it relies on user input, which is typically provided in the form of various surveys. The two types of assessment are: (a) symptom reporting, tracking, and monitoring, where users provide information about their current status in the form of valid established questionnaires, and the program assists them in monitoring and tracking their progress. Apps under this category usually send reminders for inputting symptoms at specific periods throughout the day, and then summarize the data through visual feedback to track user progress. This feature is present in 86 (52.5%) apps, and (b) self-reflection tools, which are offered through surveys and teach the users how to choose a better path, what to do in a critical situation, and other things by allowing them to explore on their own. These apps prompted users to investigate the evidence surrounding a catastrophic thinking, posed questions about the negativity in the thought process, and offered examples of how to combat the relevant negativity in their thoughts. 56 (34.4%) applications have this feature.
- **Community support:** Interacting with peers (anonymously) who are dealing with similar MH challenges or are already on the road to recovery enable users to open up about their problems and discuss viable recovery paths by exchanging support and advice. This peer-to-peer support is provided by 14 (8.8%) apps. Appointed moderators and super-admins commonly moderate these peer-to-peer discussions. A few of these apps also allow users to enlist the help of friends and family through their phone contacts and social media accounts.

|  | Google Play | % total reviews | Apple App Store | % total reviews |
|---|---|---|---|---|
| Total number of reviews | 54554 |  | 32534 |  |
| 1. After applying the length (#characters>= 200) criteria | 10485 | 19.22% | 16762 | 51.5% |
| 2. After applying the timeline (January 1, 2019 - May 1, 2021) criteria | 8556 | 15.68% | 9990 | 30.71% |
| 3a. After applying the star rating (<=3★) criteria | 1620 | 2.97% ; %reviews after applying 1 and 2: 18.93% | 2266 | 6.97%; %reviews after applying 1 and 2: 22.68% |
| 3b. Star rating (>=4★) and contain negative components | 539 | 0.98% ; %reviews after applying 1 and 2: 6.23% | 498 | 1.53%; %reviews after applying 1 and 2: 4.98% |
| Total number of reviews considered in the study: 2159 (Google Play Store); 2764 (Apple App Store) | | | | |
| Average number of Words per Review: 66.67 (Google Play Store); 113.8989 (Apple App Store) | | | | |
| Total number of apps considered after applying Filters: 117 (Google Play Store); 76 (Apple App Store) (P.S. 29 apps were available on both platforms) | | | | |

Table 1. **# of reviews from Google Play and Apple App Store after applying each inclusion criteria**

## 4.2 Perceptions and Concerns Expressed in the Critical User Reviews

Both the number of mobile MH apps and their consumers have increased significantly in recent years, which has resulted in a huge increase in the quantity of user reviews. Figure 2 depicts the recent surge in reviews across both the Android and Google platforms. Between 2016 and 2020, the number of reviews has increased by more than 20 times and 4 times in these platforms respectively. Figure 2 shows that the reviews with three or fewer stars, which we refer to as "critical reviews", accounted for about 10% of total reviews on Google Play and nearly 25% of total reviews on Apple App Store. This may not seem to be a significant percent of the userbase of these apps. However, we have to keep in mind that many of these users are MH patients who may be experiencing serious symptoms, or may be on the verge of a crisis. It is crucial to acknowledge these users' real-life experience of using the apps and take into account in planning, designing, and developing apps.

Table 1 represents a flowchart of the inclusion criteria applied and the number of reviews retrieved in each step. The final number of reviews was 2159 (3.95%, n=54554) on the Google platform and 2764 reviews (8.5%, n=32534) on the Apple platform. When it comes to writing critical reviews, users of iOS apps seem to be more comprehensive than users of Android apps (i.e., the average word count per review from Google Play is 66.67 vs. 113.89 from Apple App Store).

Our analysis identified twelve major concerns (themes) within these negative reviews. We broadly characterize these twelve themes into five overarching categories, which we report with examples from the data (themes and broad categories are listed in Table 2).

*4.2.1 Inconsistent moderation policies.* 8.8% of the apps we considered included a community feature that allowed users to share their MH problems with other users of the app and receive support. Like many online health communities (OHCs), these embedded forums and communities usually have designated admins and moderators who control destructive member posts and help answer questions to provide important support for user engagement. However, the user reviews frequently indicated that the moderation processes and policies implemented in many of the apps are quite inconsistent. Users have complaints about moderators not following established guidelines and showing biases against specific demographics, genders, races, and religions.

**Guidelines not being followed properly:** Almost all of the apps with community features have a set of guidelines in place to protect their most vulnerable users, and moderators are tasked with ensuring that these guidelines are followed. However, there have been instances where moderators have failed to follow the standard, resulting in frustrations among the users. R1 elaborated on how moderators erroneously deleted content without following app policies:

> *"Some admins have been harassing people by deleting their posts that are in the guidelines and are not bad at least. Many people have issues with the admins...They really need to regulate who they allow to be an administrator." - G0470270 (★☆☆☆☆)*

| Broad categories | Major themes | Number of reviews | Percentage of reviews |
| --- | --- | --- | --- |
| Inconsistent moderation policies | Guidelines not being followed properly | 267 | 5.4% |
| | Biased behavior from moderators and allowing discriminatory attitude | 530 | 11.2% |
| Lack of transparency | Therapist matching techniques and qualification of the therapists | 397 | 8.1% |
| | What to expect from free vs. paid subscriptions | 507 | 10.3% |
| | Ensuring privacy | 249 | 5.1% |
| Unavailability of In-time and Emergency Support | No or little help during crises | 212 | 4.3% |
| | Out-of-context responses and little interest from coaches and active listeners | 298 | 6.1% |
| Abusive community | Abuse and harassment | 409 | 8.3% |
| | Preaching harmful thoughts | 56 | 1.1% |
| Useless Chatbots and conversational agents | Not providing helpful/practical solutions | 207 | 4.2% |
| | Failing to understand context and provide personalized responses | 567 | 11.5% |
| Others (not considered) | UI issues and software bugs | 1409 | 32.3% |

Table 2. **Frequency of themes in the user reviews (number and percent of reviews expressing a particular theme, applied non-exclusively).**

Users have also expressed their dissatisfaction with the sometimes-confusing nature of these standards. Posts are often deleted and users are banned without any explanation or warning. Even when explanations are provided for deleting posts or comments, they are vague. The users felt that the moderators interpreted and enforced the norms according to their will and did not care about informing users about their actions.

At the same time, it's critical to acknowledge that maintaining rigorous rules in moderating content and banning in health peer-support communities is not always a viable option, as these decisions may result in worsening symptoms, a relapse, or even an individual's death. Some users have expressed their dissatisfaction with the moderation policies, claiming that they are too strict to express their negative emotions or to vent.

> *"For an app that's about self-expression they really do love to delete post that they don't like. This app has a stick up its a\*\*, don't waste your time." - G0670115 (★☆☆☆☆)*

It gets worse when actual offensive posts are not deleted even after reporting them. Users often complained how the moderation is *"two-faced"*. One user criticized how his/her approach of presenting a condition relating eating disorders was misconstrued and removed because the moderators lacked an understanding of particular expressions of specific MH problems. Users have reported that even after identifying and marking a large number of offensive posts by themselves, just a handful are removed by admins. Red-flagged contents, such as "threatening suicide", are left unchecked.

> *"Sometimes it can be two-faced. For example, I see posts that say 'I want to commit suicide' that never get taken down. But when I post 'I want to purge' as in I'm dealing with an eating disorder and looking for help to overcome the urge my post gets deleted. It actually makes the feeling of loneliness worse." - G0110097 (★☆☆☆☆)*

***Biased behavior from moderators and allowing discriminatory attitude:*** Users have occasionally expressed their disappointments with moderators "feeling entitled" and "abusing their power" by showing conscious biased behavior. G0470318 stated in a review that the moderators are making decisions biased towards their friends. They loosen the rules for them, and others are being unfairly punished for minor infraction of the rules. These actions place individuals in an even worse situation, as they are unable to convey, whereas biased behavior caused aggressive users to transgress the rules even more. One user compared it to biased and harmful social media groups where moderators solely target people with whom they disagree.

> *"Admins are biased, manipulative, controlling. They almost never delete the hateful posts which genuinely blow the guidelines, as those guys are friends" - G0470318 (★☆☆☆☆)*

Moderators also showed biases against different religions, races, and genders. G0470273 cited an example of moderators displaying hatred for a certain faith and allowing a violation of the guidelines in that situation.

> *"... Post bad about Islam/Muslim; it stays there, but one getting bullied & verbally abused for being a Muslim or preaching Islam, no one cares. Atleast ADMINS and MODERATORS need to show some humanity rather dealing own Islamophobia." - A04700273 (★☆☆☆☆)*

Similar to the biased and wrongful behaviors of the moderators towards specific demographics and ethnic groups, community members constantly showed racism and sexism that ruined many users' interest to share and ask for help.

> *"I have a serious problem with a toxic community when you allow people onto the platform who send you hate because of your identity or sexuality, I have mental health issues and enough problems in my life without having to worry about being told that I'm unnatural and should die." - G0470103 (★★☆☆☆)*

Another interesting finding was that, whereas usually considered minority groups experience racism, there were opposite cases as well.

> *"It is racist against whites and clearly run by Liberals. There's a Black Lives Matter section and white people are forced to silence. Admins delete the white people's comments and allow blacks to shush them." - G3010005 (★☆☆☆☆)*

Users with MH concerns may feel violated and mistreated as a result of the chaos in the moderation process, which may have unintended repercussions such as aggravating their conditions. Furthermore, such moderation behavior can make users feel uncomfortable to share, leading them to leave or uninstall the app.

*4.2.2 Lack of transparency.* Transparency in policies and services is becoming a prerequisite in many fields, particularly those where people's lives are at risk [35]. According to our analysis, there is a lack of transparency in key areas of mobile MH apps, such as how users are matched with their therapists or coaches and what their qualifications are. Users often get confused about what to expect from a paid version of the app, causing frustration after spending a handful of money. Finally, some applications do not clearly specify what user information they collect and how, infringing on users' privacy.

***Therapist matching techniques and qualification of the therapists:*** One major theme that has emerged from the user comments is how difficult it may be to navigate while looking for professional aid, such as counseling. Some apps provide unclear and incomplete instructions in the interface for searching for therapists, while some other apps completely ignore user-specified criteria in finding them a competent one. Such realities were beautifully highlighted by A1780577. The user not only complained about the app's complex interface, but also about how the application's therapy methods weren't particularly appropriate for her. It's critical for developers to retain integrity in terms of what their application offers, who their target audience is, and how the techniques/treatments they apply will benefit the consumers.

> *"I had to be experimental and ask questions to the therapist about technical aspects. A prerecorded video introduction from a potential therapist would be ideal for selecting the right one. There were short bios about a paragraph long. In hindsight, it turned out not being enough information."* - A1780577 (★★☆☆☆)

Users were concerned that the algorithms utilized in the apps to match therapists in regards to their needs were ineffective. Users recommended that the developers must pay close attention to the specifics in terms of recommending treatment procedures that are tailored to the demands of the consumers, and explain these techniques in the app in a manner that the users can understand and utilize to best meet their needs.

> *"When I told my first therapist that my family was concerned of my constant criticism & negativity, she asked if I had tried being more positive. GeE wHy didn't I tHiNk of ThAt. Really? Had I been able to be more positive, why am I paying for therapy?!?"* - A1780098 (★☆☆☆☆)

Many users suspected that the therapists they were matched with were not as professional and qualified as their in-person therapists. At times, users felt that the recommended therapists were being judgemental or dismissive which was totally unexpected to a patient suffering from a MH condition. Nonetheless, some of the users tried to find out a way around this so that they can continue using these apps.

**What to expect from free vs. paid subscriptions:** Most of the marketplace apps require users to buy a paid subscription to use full version. While users frequently called these app development companies "capitalist" and commended the developers who made the full version free during the COVID-19 pandemic, the question of whether MH apps should be free of cost or not is not the focus of our study. However, we sought to understand users' expectations from free and paid apps. Our analysis revealed that many apps provide conflicting information about what additional functionalities to expect from the paid subscription, switching to premium from a free trial, and how to request for refunds. Users occasionally downloaded and installed the apps in times of crisis to receive support, as those apps catalogued a bunch of seemingly useful features without clearly listing if and how much the users had to pay to use each feature. It often took a lot of time and effort on the user end to understand that a particular functionality in the app that attracted her is not available in the free version.

> *"I pressed a button to browse therapists and it charged me an exorbitant amount! Perhaps that is where APP is placing their bets. Exploiting people in their time of need. Nowhere on or near that button did it say that I was agreeing to pay."* - A0490010 (★☆☆☆☆)

The payment process' transparency was also challenged by users. The applications sometimes made checking the status of user payments too complex. Additionally, they were annoyed with getting charged without any prior hints.

> *"I'm a bit confused when it shows that you could have a 7 free day trial but it says that first you have to pay. How is it a free trial if u first have to pay for it?"* -G0150087 (★★★★☆)

**Ensuring privacy:** Even though certain treatment processes require sensitive information from patients in order to track and monitor symptoms, match them with interactive tools, provide better therapy and counseling services, etc., the users get wary if the apps do not provide sufficient information about what data will be collected through the app and why they will be required. For example, in G0150290, the user was terrified since their encounter with the app seemed like they were being monitored. As a patient who suffered from anxiety, it worsened their situation.

> *"When I got this app I asked it what color my shirt was, and it said the correct color. I feel watched, and scared. I thought this was supposed to relive anxiety, It made my anxiety worse." - G0150290 (★☆☆☆☆)*

A2740021 gave another example of collecting superfluous information from patients in an insensitive and inaccurate manner. The manner in which programs collect information from users is a crucial point that must be addressed.

> *"This app is always copying clipboard contents, when you open the app. This is highly illegal...Be careful. It may steal the passwords or banking info if you copied from somewhere to the clipboard!" - A2740021 (★☆☆☆☆)*

Some MH mobile applications need users to input information over an extended period of time for purposes, such as symptom monitoring, visualizing progress goals, journaling, etc. Users expect such apps to be able to securely save the information they provide and retrieve whenever needed. Additionally, in many cases the apps contained very little or no information about when and how the collected data is disseminated or shared with other stakeholders or third party. We found out that users became really skeptical of using an app if it cannot ensure the correct maintenance and security of their private information.

> *"It seems that some of the data about me and my sessions must be recorded for the app to work. But then, who can see this VERY private information about me? Can my data be sold, in aggregate or with personally identifying tags?" -G0150349 (★★☆☆☆)*

*4.2.3 Unavailability of In-time and Emergency Support.* When dealing with MH patients, it is critical to understand that a crisis can be triggered for anyone at any time, and if it is not addressed in a timely manner, it can have serious effects. A lack of concern to address this very important issue has been expressed in the reviews. After discussing a potential crisis symptom, users frequently have to wait for a long time to hear from their therapists and professional coaches, or they get short and irrelevant replies from their coaches and counselors.

**No or little help during crises:** Many times people install a MH app when in crises, and need immediate and urgent assistance (i.e., in the middle of the night when help may be unavailable, when unable to afford professional therapy, etc.). Our analysis revealed that the apps frequently failed to provide support during emergency in a timely manner. The glitches and bugs in the system often added to user frustrations.

> *" If anyone were in a state of such emergency, in need to call the suicide hotline, this app would worsen the conditions in which the struggling user would already be sick of; being ignored and feeling helpless." - A1630001 (★☆☆☆☆)*

Users also reported how usability issues with app features could worsen the crisis situation instead of helping. Almost all apps tend to have usability problems in their interface. However, the user base of these apps include people who are prone to a MH crisis and they found the usability errors interfering with their conditions.

> *"I was in a crisis and was doing a little better with the help of this app, but when I minimized to spell check something, it started me from the beginning. It made me repeat myself 3 time and sent me into a worse crisis" - G0150004 (★☆☆☆☆)*

Another crucial component in ensuring continuous support is to provide in-time responses from therapists, counselors and peers when questions asked. However, consumers have complained about long wait times and lack of follow-ups from their assigned coaches.

> *"My therapist straight up told me this is not a chat room, stop messaging me as often, I won't answer until tomorrow. I was under the impression that the app would give me help* *in a timely manner. I mean why should they give you a button for a deadline if they are going to ignore it!! Not to mention my first therapist on the app missed my deadline four times ..." - A1780031 (★☆☆☆☆)*

People lose interest in taking treatment while waiting for the prolonged wait period, which often leads to dropout. Users described the aggravation of having to wait so long for a basic response.

**Out-of-context responses and little interest from coaches and active listeners:** Despite the fact that these apps promised professional meetings with therapists and coaches, users (even after a long wait time) frequently received short, generic, out-of-context responses showing little professionalism from them. Furthermore, they (therapists and coaches) occasionally showed little interest in their patients and the therapy procedure, making consumers feel abandoned.

*"When I was assigned someone after waiting for a week, I explained my problems in a paragraph about my issues with focusing and maintaining motivation. He told me to clear my head & just study, which is completely useless because I obviously can't. When I told him that didn't help, it's not that simple, he replied with a sad-faced emoji. His 3 word response showed little interest & professionalism." - A1630001 (★☆☆☆☆)*

When working with a sensitive patient, this is a serious problem. Before selecting therapists and coaches for their applications, developers must ensure that they are well-trained and possess appropriate subject knowledge. They also need to make sure that users don't have to wait long for a response by balancing the number of users who can use the app vs the number of coaches or counselors who can help them.

*4.2.4 Abusive community.* Providing a community feature in the apps may be useful, since current research work on MH support has found online peer support to be beneficial in various contexts [80]. Users feel better after talking about their recent circumstances and concerns, and receiving advice from others who have been in similar situations. However, it is questionable how effective the community features in existing mobile MH apps have been, since we found numerous instances of toxic activities such as racism, sexism, trolling, harassment, and prejudiced that encouraged destructive ideas such as suicide.

**Abuse and harassment:** Users reported being subjected to abusive conduct and harassment. Trolling, unpleasant behavior, and harassment were all common, along with negative comments about race, gender, and religion. People with various MH concerns often use these applications as a safe place to vent and share their predicament with like-minded people, as G0470064 noted. If the atmosphere is toxic and not pleasant, users get genuinely frustrated.

*"The vast majority of comments were not only discouraging but they were absolutely rude. Very disappointed in an app that projects itself as providing help and encouragement to those who reach out for help. These humiliating and hurtful comments could actually make the individual feel a thousand times worse." - G0400064 (★☆☆☆☆)*

Users frequently blame prejudiced and harmful users who joined the community solely to take advantage of others. A few of the apps were targeted for the teenagers and for many predators it was easy to exploit these apps to abuse that population.

*"A good percentage of the people here are only there because they can get off to teenagers that don't know when they're being used. Weird questions about virginity, sex, and personal boundaries are common. Of course you could say this about any social media app, but this one is supposed to be used as a therapy alternative." - A1630032 (★☆☆☆☆)*

Oftentimes, these abusers get away with their activities by blocking the victim. On the other hand, many apps make it difficult for the users to navigate options of reporting or blocking the abusers, and even worse, some of these blocking features do not serve the purpose as expected by the app users. Blocking a user does not prevent them from harassing others; in some applications, they can still harass others by commenting on other posts, checking in on the victims, and so on.

> *"If someone blocks you you cannot report them, this feature on an app filled with sad, lonely, socially unaware individuals is a terrible idea. It's like the app was made for abusers to use it however they want, no negative repercussions." - A1530032 (★★☆☆☆)*

**Preaching harmful thoughts:** When the environment becomes too unhealthy, it might lead to detrimental thoughts, such as advocating suicide. G0470163 witnessed one such terrible situation in which, rather than offering intelligent advice and consolation, friends made cruel comments and pushed the victim to continue down the path of destruction.

> *"The app has turned into more of a popularity contest and toxic community. I read a comment stating 'Please do it' on someone's post who said they wanted to die. The community has gone down the drain, it gives the opposite aspect this app originally tried to deliver; a safe place to express your feelings without any fear of trolling." - G0470163 (★☆☆☆☆)*

Peers often encouraged unhealthy lifestyles and harmful thoughts, such as "suicidal ideation," "self harm," and "dangerous self medicating."

> *"I've seen people be absolutely ignorant in the comments of people posting about their struggles, even worse, people blindly encouraging others to become more unhealthy when it's already dangerous." - G040068" (★☆☆☆☆)*

Providing a safe environment where users can unwind and communicate to one another should be a key concern for programs that provide peer-to-peer support, many of these apps failed to offer this after projecting themselves as such. This is also directly related to the preceding theme in Section 4.2.1, which indicates that the moderation procedure isn't being carried out effectively. To provide a safe atmosphere, consistent and effective moderation is required.

*4.2.5 Useless Chatbots and conversational agents.* Researchers have proposed several approaches to employ artificial intelligence (AI) and machine learning algorithms in improving lifestyles and decision support systems in recent years, thanks to new innovation in AI algorithms. One method is to include these algorithms into a bot that can converse with people who have MH issues. These are known as 'Chatbots'. This theme addresses concerns relating to many aspects of chatbots, such as failure to be helpful, lack of learning capabilities, and risk of triggering MH issues through incorrect responses.

**Not providing helpful/practical solutions:** Chatbots assist patients with mental illnesses in a variety of ways, such as by interacting with them, providing comfort through intelligent responses, recommending healthier fitness and lifestyle choices. They are efficient in some respects, but have been criticized for being unhelpful. They have occasionally failed to respond to concerns in a responsible way, and have also been found to be ineffective. Understanding the situations is critical in the rehabilitation process, and these bots have occasionally failed to perform what they were designed to do due to poor algorithms and construction. G0150093 mentioned such a case where AI was poorly constructed and failed to understand and address the issue.

> *"Feels like more of a resource finder as it could not understand what I was trying to ask it. The Ai wanted to ask all the questions and set me on a certain path. I hope that Ai will advance in the future and be able to help more. Also it told me I needed to sleep when I didn't and kept telling me how it would help my energy levels, when in fact I'm almost always tired and sleep rarely helps." - G0150093 (★☆☆☆☆)*

Furthermore, understanding and mimicking how an actual person responds to questions is crucial. Word choice and voice tone are vital when interacting with users experiencing MH concerns, since all these can have an impact on the outcome of the interaction. User reviews demonstrated how bots often conversed as if users were pets, and undermining the positivity that the bot was attempting to convey. In addition, dismissive behaviors from AI agents were noticed frequently.

> *"This morning I received the daily text, addressing me as a dog owner would: 'Who's a good user? You areeeee. Yes you are! Your Monday motivation hack: Talk to yourself like a pet.' Seriously? This is demeaning in its endeavor to encourage patronizing tones on oneself rather than truly home an empowering message. Speaking the language of dumb and basic was never considered a strategy for self-respect which makes this even more asinine. ..." - A1660001 (★☆☆☆☆)*

> *"It cut me off & said goodbye before I was finished. For someone who is emotionally unstable I imagine even a robot dismissing u to be a blow u don't need." -G0150222 (★☆☆☆☆)*

Feelings and emotions can be difficult for the bot to comprehend at times. The ability to handle an emergency situation is a vital trait that these AI agents must possess. In other programs, users have expressed dissatisfaction with how they respond in those scenarios. Poor usability and over complication were mentioned by some users. Even though users realize that bots can be occasionally helpful, AI ineffectiveness frustrates users more.

***Failing to understand context and provide personalized responses:*** Users reported having "incredibly boring conversations with the bots as AI misses a lot of contextual clues" (G0150164) or "failed to pick an appropriate reaction for the context of the conversation".

> *"BOT very quickly starts to feel exactly like that: a bot. It loses context, chats start to make less sense. Some things it says are very out of place. It feels like it's ignoring you and is expecting a response that you obviously didn't give" - A1660034 (★☆☆☆☆)*

Some users were more irritated. While they understand it is not an actual human, and that, talking to it made them feel more at ease, it was at times of no help as it could not answer basic inquiries and kept repeating replies. Users have been led to assume that the bots don't comprehend what they're saying because they respond with incredibly irrelevant or repetitive responses. G0150304 demonstrated how frustrating these responses may be, and it eventually made the user exit the applications.

> *"I was trying to tell BOT that I'm really glad about my day and I told BOT why, then it keeps asking me 'Why do you say so' for about 3 times like what the?! I feel like it's the bot who needs help, not me." - G0150304 (★☆☆☆☆)*

Furthermore, consumers expressed dissatisfaction with how generic some of the responses and recommended activities were. They reported of being "offered more generic suggestions than they could find from simple google searches." It became worse when the bot suggested activities that the user was unable to perform due to physical or ambulatory reasons. In general, the reviews emphasized that mental health conditions can cause complex and varied moods among users that are hard to comprehend, and that the users expected these apps to incorporate features adaptive to these feelings and moods.

> *"Bot doesn't understand simple keywords like 'physical pain' or 'headache.' Offers more generic suggestions than I have found on simple google searches. ..." - G0150203 (★★☆☆☆)*

> *"I just have one suggestion: in worry relievers add more moods to it because sometimes what you actually is feeling inside is not as simple as anger,happy ,sad and okay. Like I usually feel nothing at all or so numb and that is the situation that i want to deal with the most on daily basis." - G0150277 (★★★★☆)*

Some bots occasionally asked "preset questions" that did not apply to many, "steamrolled users into a limited range of non-useful and non-personalized responses," acted "judgmental rather than self-affirming," advocated "only a single line of thought."

> *"I expected an AI but it seems preset questions, though questions are well placed. The problem lies when the human deviates from the line of talk, begins rambling. Then it gradually fails to understand us, because the questions are preset." - A2640093 (★★☆☆☆)*

In one case, a bot with the intention of reforming a user's negative thoughts erased and modified part of his journal that frustrated him and he started questioning the reliability of the the app. This is a practical example of how an intervention strategy that's been supported by the literature to improve particular MH conditions can backfire, and thus illustrates the significance of understanding the consequences that can result from adding a feature.

> *"APP censors your journal entries. what kind of journal CHANGES what you wrote?! It asked me my thoughts, I started waxing poetic metaphors about my problems. It censored it so I don't even know what I was talking about anymore! Some of my conversations are completely lost, they don't show in the journal." - A1630013 (★★★★☆)*

### 4.3 Aspects and features of MH apps liked by the users

Due to their convenience and accessibility, mobile mental health apps have proven to be beneficial. Despite the fact that we conducted thematic analysis on the negative contents of all reviews that met our filter parameters, we evaluated all of the positive reviews after applying filter 1 and filter 2 *(Table 1)* as it's also crucial to understand user preferences of different features of these apps. We point out a few of the significant themes from the positive reviews, that we think are necessary to consider while implementing MH apps and developing guidelines for these apps:

*4.3.1 Easy navigation.* : Users appreciated the apps with simple and easy flow of navigation. Low cognitive skill is preferred by those with various mental health difficulties such as worry, or stress, according to the users. They loved "steps that gradually walk them through techniques they learn in therapy," "design or amount of information that they don't feel overwhelmed with," and "the good little tips".

*4.3.2 Actionable and practical features.* : Users positively acknowledged the activities and exercises that are ''bite-sized", consume low attention span, and take short time to complete. Reviews positively mentioned the visual and interactive modules that put theory in action.

> *"Great step by step daily instruction. I've read a lot of CBT workbooks and this app puts them into motion. As a highly visual learner, the videos in the app are huge in helping me to remember and apply techniques." - G0260019 ★★★★★*

> *"Really enjoying the actionable steps in this app. The stages of each journey don't take too long. Everything is bite-sized and helpful." - A0390002 ★★★★★*

*4.3.3 Higher benefit-cost ratio.* : Most apps do not allow users complete access if they do not pay for the subscription. A handful of reviews revealed that they compare these apps with alternative options, determine the benefit-cost ratio for them, and do not mind paying if they think the ratio is considerably high, especially for apps that support treatment through expert therapists, guiding coaches, and active listeners.

> *"I've tried face-to-face therapy before. I always feel like I have to perform. Writing is so much more therapeutic for me. I am ok with paying the subscription and have so far had one session with my coach which was excellent." - G0120007 ★★★★★*

Many people prefer having therapy and coaching sessions via mobile devices because it is more convenient. Furthermore, many customers had no problems paying because these app-delivered therapies were covered by their health insurance. There was an abundance of positive comments about how they didn't have to worry about the payments because they were taken care of by insurance.

> *"My insurance covers the cost of the upgraded version. Thank you and bless you to the masterminds staffing this app. 'Providing hope' to others is a gift!" - A2770005* ★★★★★

*4.3.4 Chatbots as companions.* : Chatbots are a specific tool that, according to users, can have both positive and negative repercussions. One of the major benefits we have observed from user reviews is that, chatbots can makes them feel less lonely. Many users noted that now that they have chatbots in their smartphone apps, they have someone to talk to 24 hours a day, seven days a week.

> *"It's like you're talking to a friend but it's really more than a friend and really, REALLY GREAT FRIEND! You'll realize that this app is much chatting with a such very sweet and cute friendly *animation that will help you with your problems!" - G1120063* ★★★★★

One of the most notable distinctions between talking to a person and interacting with chatbots has been the bots' lack of judgment. People do not anticipate AI acting in the role of a professional therapist. Rather, they remarked on how effective it was at directing their good thoughts and making them feel better. Many users were concerned about which pronouns (he/she/them) they should use for chatbots, proving the fact that it had been acknowledged as a distinct entity to them.

> *"BOT helps me to change my manner of thinking within minutes. From being pessimistic and hopeless to feeling confident and hopeful. I check in a few times a day depending on when I feel myself slipping back to old ways. I believe that over the long term this app will help me in all aspects of my life exponentially. It's like you're talking to a friend who really knows how you're feeling and cares." - G0870009* ★★★★★

## 5 DISCUSSION

In this section, we bring together insights from the findings to provide recommendations and future directions for mobile MH app design.

### 5.1 Research and Design Implications for MH Apps Development

*5.1.1 Towards consistent moderation and community support.* Even though careful and considerate moderation is an essential aspect of developing a community support system [120], our analysis demonstrated how poorly the app-embedded communities are moderated (i.e., moderators showing biased behavior against specific races, genders, and other marginalized communities, allowing harassment of members on multiple occasions, etc.). Inconsistent and careless moderating decisions may allow abusive community behavior such as trolling, harassment, and discrimination to flourish, severely impacting users and triggering dangerous conditions in this sensitive context (*section 3.2.4*). These findings have a major bearing on the necessity of developing and following specific moderation policies curtailed to the community needs. Following directions from prior studies can be effective, such as appointing moderators with well-defined credentials to build trust in OHCs (i.e., if the moderators are health professionals vs. with no clinical background) [50, 53] or relying on indirect forms of social control to avoid punishing kinds of control [67]. Because mobile MH communities are significantly smaller in scale than OHCs, future research should build on these findings and their implications on a smaller scale platform by gaining an understanding of the underlying mechanisms that explain how different styles of moderation work. Some small-scale online communities may not have enough resources and capacities to appoint dedicated moderators and regulate a continuous moderation process, peer-led moderation can be considered as an alternative there. Peer-led moderation can support peers by offering non-medical guidance, and plays a crucial role in community involvement [98, 102].

Moderators often become less engaged as they feel overwhelmed to manually (dis)approve large volume of content [99]. Future app designers may consider employing natural language processing techniques to enhance the pre-moderation stage and flag content automatically for human review. However, there are various concerns and challenges with AI content moderation, which can be addressed by including humans in the feedback loop and boosting the model's transparency [44, 94, 95]. Automated techniques can be employed for continuous assessment of moderation decisions to mark racist moderation behaviors.

*5.1.2 Towards better transparency.* Many users were genuinely concerned and puzzled about what information is being collected and how. Whereas, data transparency is critical to gaining patients' trust [54] in health-related systems, rules, regulations, and internal policies on information sharing have made it difficult to maintain complete transparency [58, 104]. To strike a balance between what users desire and how much information third parties are prepared to offer, researchers should focus on collaboration between multiple stakeholders - users, providers, and developers and examine their values and needs in terms of data transparency [64].

Users criticized that they download an app when already feeling low, just to realize that the support they need is available through the app, but they cannot afford it. Additionally, apps frequently failed to provide a free trial and instead charge through hidden features in order to make profit, or even to collect the maintenance cost of the apps. Users expressed that a free trial would be convenient before paying for something that might not work for their specific needs. We understand that there are costs associated with developing and maintaining the apps. However, developers should carefully design the interfaces that make pricing information transparent. For instance, they can adopt simple visualization approaches that show the pricing breakdown and total cost upfront and provide reminders and warning notifications before any type of payment is charged. Moreover, free trials should be offered for the useful features, and if that's not the case, then inform users even before they attempt to install the app that none of the services offered are free. According to our assessment of positive reviews, if users find the tools satisfactory during the free trial, they are willing to pay for them afterward. Some people expressed their delight that their insurance covers the cost of the subscription. However, mobile mental health technology is seen as a convenient and accessible tool for all people, regardless of their capacity to pay for health insurance. As a result, developers and policymakers should develop an economical solution for both individuals who have health insurance and those who do not.

*5.1.3 Towards usability and UX.* Many users installed the applications to avoid and identify any unexpected crises, whereas the apps' inability to meet expectations worsened their mental health conditions. Users were perplexed about how they were matched with their therapists and counselors, since the assigned ones often did not communicate as expected. It's fair if the initial therapist someone is matched with is not right for them, but user complaints about numerous iterations of failed sessions with different therapists are very concerning. Our findings suggest users appreciate easy navigation, detailed explanation of how the tools work and motivational contents (such as success stories). A simple interactive checklist interface can be used to take user symptoms and other preferences as input and recommendation algorithms can be adopted to show them an ordered list of therapists based on their symptoms. Moreover, therapists should be offered with interfaces to easily list their qualifications, as well as their previous history of dealing with patients in anonymized case story format. Prior research suggests that it may bring more credibility to the users [34] and boost user satisfaction with the therapy process [26].

*5.1.4 Towards app-embedded chatbot design.* Users feel safe to share health conditions with a conversational agent over a real person [1], so chatbots should be utilized to its full potential to enhance mental health support for a broad range of users. User reviews revealed a number of shortcomings in existing chatbots, such as providing unhelpful and repetitive responses making the interaction unproductive, and being dismissive or patronizing that made users feel neglected and disrupted the treatment efforts. Our findings suggested that users have a preference for personalized responses, and conversations that normally draw on CBT techniques to assist users address any harmful thought and behavior patterns and build coping mechanisms must be more contextual and relevant, similar to prior work [17]. Different methodologies for optimizing chatbot responses [16], including utilizing linguistic aspects, such as employing emotion or sentiment analysis [78] have been proved effective for better product suggestions and providing customer care. These implementations and approaches can be customized for chatbots in MH apps to more engaging and capable of displaying positive attitudes. Alternatively, if the bot isn't AI-powered, developers should make it clear to consumers that it's merely a bot with pre-programmed responses.

It is crucial to consider that the targeted participants may be going through different challenges and a crisis can happen anytime. The existing chatbots are unable to identify potential crises and provide effective options, as depicted in our analysis. Existing machine learning and pattern detection algorithms can be employed to detect potential crises from users' interaction with the chatbots and take appropriate action by alerting their healthcare providers or friends and family members [115]. Our findings suggest following participatory machine learning approaches to design technology for vulnerable populations that can foster interactions among different stakeholders of these technologies. This technique will seem useful to academics for future research, as well as to developers and health care practitioners to implement technology-driven therapies.

## 5.2 Need for Developing a User-Centric Guideline for MH Apps Design

Although there are several guidelines available for developers of MH apps (Section 2.3), none of them are comprehensive and they are dispersed across various organizational resources. No significant effort has been made to disseminate these policies, either [86]. It is an unrealistic expectation that developers are able to search for and follow proper guidelines and recommendations on their own, especially if they have no prior experience. We provide suggestions (informed by our findings) on how app developers' and organizations' access to existing guidelines and regulations can be facilitated, and for creating a comprehensive centralized guideline for them.

The Federal Trade Commission (FTC) has developed a tool that advises developers on which federal rules and regulations they should adhere to, but only in the legal sense [37]. There are other useful recommendations available established by various professional and medical organizations, such as expert guidelines from the American Psychiatric Association (APA) [4], non-profit organization psyberguide [81], etc. Bringing all these information together in one place via an interactive tool can help developers get aware and act more carefully before beginning to development an app. Such tools may prompt developers to select different aspects of the app they are planning to develop (e.g., targeted users, techniques and tools to be used, etc.) and show appropriate regulations and guidelines to follow based on their selections. These tools can potentially be disseminated through incorporating them in common Integrated Development Environments (IDEs), such as Xcode [55] and Android studio [31].

It can be deduced from the numerous and repeated user criticisms that their feedback is ignored even after many updates of the apps, making us believe that there is a communication gap between the users and the developers. To promote accountability, developers may occasionally be provided with snapshots of negative reviews as well as the health consequences stemming from those. Many available apps are not evidence-based [8], and are only tangentially informed by evidence-based framework, but claim to provide an essential support system and treatment to a wide range of people suffering from a variety of mental health conditions [8, 32]. It is critical that developers examine the efficacy of their enforced approaches on a small population before making it public.

According to our findings in this article, a comprehensive guideline should consider at least the following elements. Note that this is not a complete guideline and to implement an effective guideline more research and analysis of MH app data from other sources are necessary.

- If an app aims to provide peer-to-peer help, developers need to be reminded to implement techniques that keep the moderators and admins accountable. If volunteer moderators are being appointed in a community, a list of persuasive strategies might be shown to make them aware of their responsibilities and to be respectful of the policies.
- Users appreciate well-explained information regarding the support they would be provided with. Transparency is crucial in this context, and guidelines for development should show tips around enhancing transparency in all components of an app (e.g., algorithms/inner mechanisms, subscription, privacy of information, etc.). One of the simplest techniques in this regard could be inclusion of an interactive interface in the guideline that developers can use to design a customized FAQ page for their apps.
- Apps often collect private information from users to recommend appropriate and personalized treatment. Because there are so many guidelines for collecting, maintaining, and using data, it's impossible for developers to figure out which ones apply to their specific situation. A comprehensive guideline should keep track of the latest updates of these regulations and instruct the developers in an easy to follow manner.
- One of the vulnerabilities of these MH app users is the likelihood of their condition being triggered or worsened. A successful guideline should contain automated approaches to predict such consequences from a feature being developed. It can mine and summarize user and expert reviews from multiple sources and show warnings throughout the app development.

Certain issues caused users to abandon the app and the recovery process. It is important to keep in mind that most (if not all) developers are not medical professionals. Some troubles faced by the users may appear insignificant to them, that have serious health consequences. A variety of participatory research and design techniques [11], including workshops [65], focus groups [43], ethnography [12], etc. can be replicated on a relatively small scale to facilitate conversations among different stakeholders. This should as well be followed for continuous evaluation of the applications' efficacy, to get guidance on what to do and what to avoid for the next update.

## 6 CONCLUSION

There is a gap in the current literature in terms of recognizing the challenges faced by the MH app users that affect adherence and engagement. First, to gain an empirical understanding of the current state of mobile app marketplaces we examine the descriptions of 164 mental health apps from Google Play and Apple App Stores. Next, we conducted an inductive thematic analysis of 1620 Android user reviews and 2266 iOS user reviews to better understand usage patterns, their perspectives and concerns. Our findings include criticisms of inconsistent moderation standards, a lack of transparency in how people were matched with their therapists/ counselors, as well as what qualifications they have, a lack of transparency in what customers get in the free vs paid version, and how user data privacy is protected. We also noticed a lack of real-time and emergency assistance and the ineffective implementations of app-embedded chatbots. This study's contributions include research and design implications for consistent moderating and community support, as well as improved transparency, usability, and UX. We also offered a few suggestions regarding the creation, implementation, and dissemination of a comprehensive guideline for MH app developers.

## 7 ACKNOWLEDGEMENTS

This work is partially supported by the Northwestern Mutual Data Science Institute (01810-43826).

## REFERENCES

[1] Alaa A Abd-Alrazaq, Mohannad Alajlani, Nashva Ali, Kerstin Denecke, Bridgette M Bewick, and Mowafa Househ. 2021. Perceptions and opinions of patients about mental health chatbots: Scoping review. *Journal of medical Internet research* 23, 1 (2021), e17828.

[2] Felwah Alqahtani and Rita Orji. 2019. Usability issues in mental health applications. In *Adjunct Publication of the 27th Conference on User Modeling, Adaptation and Personalization*. 343–348.

[3] Felwah Alqahtani and Rita Orji. 2020. Insights from user reviews to improve mental health apps. *Health informatics journal* 26, 3 (2020), 2042–2066.

[4] title =App Advisor url =https://www.psychiatry.org/psychiatrists/practice/mental-health-apps lastaccessed =August 25, 2021 American Psychiatric Association (APA), year = 2021. [n.d.]. .

[5] title =App Store Review Guidelines url =https://developer.apple.com/app-store/review/ lastaccessed =August 25, 2021 Apple Inc., year = 2017. [n.d.]. .

[6] Patricia A Arean, Kevin A Hallgren, Joshua T Jordan, Adam Gazzaley, David C Atkins, Patrick J Heagerty, and Joaquin A Anguera. 2016. The use and effectiveness of mobile apps for depression: results from a fully remote clinical trial. *Journal of medical Internet research* 18, 12 (2016), e330.

[7] GSM Association et al. 2012. Gsma mobile and privacy, privacy design guidelines for mobile application development.

[8] David Bakker, Nikolaos Kazantzis, Debra Rickwood, and Nikki Rickard. 2016. Mental health smartphone apps: review and evidence-based recommendations for future developments. *JMIR mental health* 3, 1 (2016), e4984.

[9] Michael Bauer, Tasha Glenn, John Geddes, Michael Gitlin, Paul Grof, Lars V Kessing, Scott Monteith, Maria Faurholt-Jepsen, Emanuel Severus, and Peter C Whybrow. 2020. Smartphones in mental health: a critical review of background issues, current status and future concerns. *International journal of bipolar disorders* 8, 1 (2020), 1–19.

[10] Aashutosh Bhatt, Ankit Patel, Harsh Chheda, and Kiran Gawande. 2015. Amazon review classification and sentiment analysis. *International Journal of Computer Science and Information Technologies* 6, 6 (2015), 5107–5110.

[11] Erling Björgvinsson, Pelle Ehn, and Per-Anders Hillgren. 2012. Design things and design thinking: Contemporary participatory design challenges. *Design issues* 28, 3 (2012), 101–116.

[12] Design Jeanette Blomberg and Helena Karasti. 2012. Ethnography: Positioning ethnography within participatory design. In *Routledge international handbook of participatory design*. Routledge, 106–136.

[13] Sandra Bucci, Natalie Berry, Rohan Morris, Katherine Berry, Gillian Haddock, Shôn Lewis, and Dawn Edge. 2019. "They Are Not Hard-to-Reach Clients. We Have Just Got Hard-to-Reach Services." Staff Views of Digital Health Tools in Specialist Mental Health Services. *Frontiers in psychiatry* 10 (2019), 344.

[14] Jane Burns. 2013. Game on: Exploring the impact of technologies on young men's mental health and wellbeing. (2013).

[15] Kim Burns, Ranmalie Jayasinha, and Henry Brodaty. 2017. Evaluation of an electronic app developed to assist clinicians in the management of behavioral and psychological symptoms of dementia (BPSD). *International Journal of Human–Computer Interaction* 33, 11 (2017), 902–910.

[16] Gillian Cameron, David Cameron, Gavin Megaw, Raymond Bond, Maurice Mulvenna, Siobhan O'Neill, Cherie Armour, and Michael McTear. 2017. Towards a chatbot for digital counselling. In *Proceedings of the 31st International BCS Human Computer Interaction Conference (HCI 2017) 31*. 1–7.

[17] Gillian Cameron, David Cameron, Gavin Megaw, Raymond Bond, Maurice Mulvenna, Siobhan O'Neill, Cherie Armour, and Michael McTear. 2018. Assessing the usability of a chatbot for mental health care. In *International Conference on Internet Science*. Springer, 121–132.

[18] Claire Cardie, Janyce Wiebe, Theresa Wilson, and Diane J Litman. 2003. Combining Low-Level and Summary Representations of Opinions for Multi-Perspective Question Answering.. In *New directions in question answering*. 20–27.

[19] Andrew D Carlo, Reza Hosseini Ghomi, Brenna N Renn, and Patricia A Areán. 2019. By the numbers: ratings and utilization of behavioral health mobile applications. *NPJ digital medicine* 2, 1 (2019), 1–8.

[20] Steven Chan, John Torous, Ladson Hinton, and Peter Yellowlees. 2015. Towards a framework for evaluating mobile mental health apps. *Telemedicine and e-Health* 21, 12 (2015), 1038–1041.

[21] Judith A Chevalier and Dina Mayzlin. 2006. The effect of word of mouth on sales: Online book reviews. *Journal of marketing research* 43, 3 (2006), 345–354.

[22] Tya Chuanromanee and Ronald Metoyer. 2020. Evaluation and Comparison of Four Mobile Breathing Training Visualizations. In *2020 IEEE International Conference on Healthcare Informatics (ICHI)*. IEEE, 1–12.

[23] Eric K Clemons, Guodong Gordon Gao, and Lorin M Hitt. 2006. When online reviews meet hyperdifferentiation: A study of the craft beer industry. *Journal of management information systems* 23, 2 (2006), 149–171.

[24] Federal Trade Commission et al. 2015. FTC charges marketers of 'vision improvement'app with deceptive claims. *USA: https://www. ftc. gov/news-events/press-releases/2015/09/ftc-chargesmarketers-vision-improvement-app-deceptive-claims* (2015).

[25] Federal Trade Commission et al. 2015. FTC cracks down on marketers of "Melanoma Detection" apps. *Retrieved on February* 1 (2015), 2016.

[26] Kathryn E Cox, Laura M Simonds, and Alesia Moulton-Perkins. 2021. Therapist-targeted googling: Characteristics and consequences for the therapeutic relationship. *Professional Psychology: Research and Practice* (2021).

[27] title =Aspects to Consider When Developing a Mental Health App url =https://yalantis.com/blog/mental-health-app-development/ lastaccessed =August 25, 2021 Daria Bulatovych, year = 2021. [n.d.]. .

[28] Fernando Estrada Martinez de Alva, Greg Wadley, and Reeva Lederman. 2015. It feels different from real life: users' opinions of mobile applications for mental health. In *Proceedings of the annual meeting of the Australian special interest group for computer human interaction.* 598–602.

[29] Anjali Devakumar, Jay Modh, Bahador Saket, Eric PS Baumer, and Munmun De Choudhury. 2021. A Review on Strategies for Data Collection, Reflection, and Communication in Eating Disorder Apps. In *Proceedings of the 2021 CHI Conference on Human Factors in Computing Systems.* 1–19.

[30] title =Application of self-Regulatory Principles to the Mobile environment url =shorturl.at/kxCJ6 lastaccessed =August 25, 2021 DIGITAL ADVERTISING ALLIANCE, year = 2013. [n.d.]. .

[31] Jerome DiMarzio. 2016. *Beginning Android Programming with Android Studio.* John Wiley & Sons.

[32] Tara Donker, Katherine Petrie, Judy Proudfoot, Janine Clarke, Mary-Rose Birch, and Helen Christensen. 2013. Smartphones for smarter delivery of mental health programs: a systematic review. *Journal of medical Internet research* 15, 11 (2013), e2791.

[33] Wenjing Duan, Qing Cao, Yang Yu, and Stuart Levy. 2013. Mining online user-generated content: using sentiment analysis technique to study hotel service quality. In *2013 46th Hawaii International Conference on System Sciences.* IEEE, 3119–3128.

[34] Thomas Anthony Dyer, Janine Owens, and Peter Glenn Robinson. 2014. The acceptability of care delegation in skill-mix: the salience of trust. *Health Policy* 117, 2 (2014), 170–178.

[35] Motahhare Eslami, Kristen Vaccaro, Min Kyung Lee, Amit Elazari Bar On, Eric Gilbert, and Karrie Karahalios. 2019. User attitudes towards algorithmic opacity and transparency in online reviewing platforms. In *Proceedings of the 2019 CHI Conference on Human Factors in Computing Systems.* 1–14.

[36] Maria Faurholt-Jepsen, Mads Frost, Christian Ritz, Ellen Margrethe Christensen, AS Jacoby, Rie Lambæk Mikkelsen, U Knorr, JE Bardram, Maj Vinberg, and Lars Vedel Kessing. 2015. Daily electronic self-monitoring in bipolar disorder using smartphones–the MONARCA I trial: a randomized, placebo-controlled, single-blind, parallel group trial. *Psychological medicine* 45, 13 (2015), 2691–2704.

[37] title =MOBILE HEALTH APPS INTERACTIVE TOOL url =https://www.ftc.gov/tips-advice/business-center/guidance/mobile-health-apps-interactive-tool lastaccessed =August 25, 2021 Federal Trade Commission, year = 2016. [n.d.]. .

[38] Geraldine Fitzpatrick and Gunnar Ellingsen. 2013. A review of 25 years of CSCW research in healthcare: contributions, challenges and future agendas. *Computer Supported Cooperative Work (CSCW)* 22, 4 (2013), 609–665.

[39] Theresa Fleming, Lynda Bavin, Mathijs Lucassen, Karolina Stasiak, Sarah Hopkins, Sally Merry, et al. 2018. Beyond the trial: systematic review of real-world uptake and engagement with digital self-help interventions for depression, low mood, or anxiety. *Journal of medical Internet research* 20, 6 (2018), e9275.

[40] Chris Forman, Anindya Ghose, and Batia Wiesenfeld. 2008. Examining the relationship between reviews and sales: The role of reviewer identity disclosure in electronic markets. *Information systems research* 19, 3 (2008), 291–313.

[41] Karen Fortuna, Paul Barr, Carly Goldstein, Robert Walker, LaPrincess Brewer, Alexandra Zagaria, and Stephen Bartels. 2019. Application of community-engaged research to inform the development and implementation of a peer-delivered mobile health intervention for adults with serious mental illness. *Journal of participatory medicine* 11, 1 (2019), e12380.

[42] Bin Fu, Jialiu Lin, Lei Li, Christos Faloutsos, Jason Hong, and Norman Sadeh. 2013. Why people hate your app: Making sense of user feedback in a mobile app store. In *Proceedings of the 19th ACM SIGKDD international conference on Knowledge discovery and data mining.* 1276–1284.

[43] Jonas Geuens, Luc Geurts, Thijs W Swinnen, René Westhovens, Maarten Van Mechelen, and Vero Vanden Abeele. 2018. Turning tables: A structured focus group method to remediate unequal power during participatory design in health care. In *Proceedings of the 15th Participatory Design Conference: Short Papers, Situated Actions, Workshops and Tutorial-Volume 2.* 1–5.

421:26 MD Romael Haque and Sabirat Rubya

[44] Robert Gorwa, Reuben Binns, and Christian Katzenbach. 2020. Algorithmic content moderation: Technical and political challenges in the automation of platform governance. *Big Data & Society* 7, 1 (2020), 2053951719897945.

[45] Jennifer Green, Jennifer Huberty, Megan Puzia, Chad Stecher, et al. 2021. The Effect of Meditation and Physical Activity on the Mental Health Impact of COVID-19–Related Stress and Attention to News Among Mobile App Users in the United States: Cross-sectional Survey. *JMIR mental health* 8, 4 (2021), e28479.

[46] Emitza Guzman, Luís Oliveira, Yves Steiner, Laura C Wagner, and Martin Glinz. 2018. User feedback in the app store: a cross-cultural study. In *2018 IEEE/ACM 40th International Conference on Software Engineering: Software Engineering in Society (ICSE-SEIS)*. IEEE, 13–22.

[47] Virginia Harrison, Judith Proudfoot, Pang Ping Wee, Gordon Parker, Dusan Hadzi Pavlovic, and Vijaya Manicavasagar. 2011. Mobile mental health: review of the emerging field and proof of concept study. *Journal of mental health* 20, 6 (2011), 509–524.

[48] Daojing He, Menghan Pan, Kai Hong, Yao Cheng, Sammy Chan, Xiaowen Liu, and Nadra Guizani. 2020. Fake Review Detection Based on PU Learning and Behavior Density. *IEEE Network* 34, 4 (2020), 298–303.

[49] Minqing Hu and Bing Liu. 2004. Mining and Summarizing Customer Reviews. In *Proceedings of the Tenth ACM SIGKDD International Conference on Knowledge Discovery and Data Mining* (Seattle, WA, USA) *(KDD '04)*. Association for Computing Machinery, New York, NY, USA, 168–177. https://doi.org/10.1145/1014052.1014073

[50] Jina Huh, Rebecca Marmor, and Xiaoqian Jiang. 2016. Lessons learned for online health community moderator roles: a mixed-methods study of moderators resigning from WebMD communities. *Journal of medical Internet research* 18, 9 (2016), e247.

[51] Yavuz Inal, Jo Dugstad Wake, Frode Guribye, Tine Nordgreen, et al. 2020. Usability evaluations of mobile mental health technologies: systematic review. *Journal of medical Internet research* 22, 1 (2020), e15337.

[52] Ahmad Kamal. 2013. Subjectivity classification using machine learning techniques for mining feature-opinion pairs from web opinion sources. *arXiv preprint arXiv:1312.6962* (2013).

[53] Shaheen Kanthawala and Wei Peng. 2021. Credibility in Online Health Communities: Effects of Moderator Credentials and Endorsement Cues. *Journalism and Media* 2, 3 (2021), 379–396.

[54] Bonnie Kaplan. 2020. Seeing through health information technology: the need for transparency in software, algorithms, data privacy, and regulation. *Journal of Law and the Biosciences* 7, 1 (2020), lsaa062.

[55] Maurice Kelly and Joshua Nozzi. 2013. *Mastering Xcode: Develop and Design.* Peachpit Press.

[56] Michelle E Kiger and Lara Varpio. 2020. Thematic analysis of qualitative data: AMEE Guide No. 131. *Medical teacher* 42, 8 (2020), 846–854.

[57] Darrell G Kirch. 2020. Physician mental health: My personal journey and professional plea. *Academic Medicine* 96, 5 (2020), 618–620.

[58] Patty Kostkova, Helen Brewer, Simon de Lusignan, Edward Fottrell, Ben Goldacre, Graham Hart, Phil Koczan, Peter Knight, Corinne Marsolier, Rachel A McKendry, et al. 2016. Who owns the data? Open data for healthcare. *Frontiers in public health* 4 (2016), 7.

[59] SJ Koyani et al. 2004. Research-based web design & usability guidelines (p. 232). *National Cancer Institute* (2004).

[60] Nanda Kumar and Izak Benbasat. 2006. Research note: the influence of recommendations and consumer reviews on evaluations of websites. *Information Systems Research* 17, 4 (2006), 425–439.

[61] Wooseok Kwon, Minwoo Lee, and Ki-Joon Back. 2020. Exploring the underlying factors of customer value in restaurants: A machine learning approach. *International Journal of Hospitality Management* 91 (2020), 102643.

[62] Jianwei Lai, Peng He, Hsien-Ming Chou, and Lina Zhou. 2013. Impact of national culture on online consumer review behavior. *Global Journal of Business Research* 7, 1 (2013), 109–115.

[63] Emily G Lattie, Eleanor Burgess, David C Mohr, and Madhu Reddy. 2021. Care Managers and Role Ambiguity: The Challenges of Supporting the Mental Health Needs of Patients with Chronic Conditions. *Computer Supported Cooperative Work (CSCW)* 30, 1 (2021), 1–34.

[64] Karen Luxford, Dana Gelb Safran, and Tom Delbanco. 2011. Promoting patient-centered care: a qualitative study of facilitators and barriers in healthcare organizations with a reputation for improving the patient experience. *International Journal for Quality in Health Care* 23, 5 (2011), 510–515.

[65] Ezio Manzini and Francesca Rizzo. 2011. Small projects/large changes: Participatory design as an open participated process. *CoDesign* 7, 3-4 (2011), 199–215.

[66] Daniel Martens and Walid Maalej. 2019. Towards understanding and detecting fake reviews in app stores. *Empirical Software Engineering* 24, 6 (2019), 3316–3355.

[67] Uwe Matzat and Gerrit Rooks. 2014. Styles of moderation in online health and support communities: An experimental comparison of their acceptance and effectiveness. *Computers in Human Behavior* 36 (2014), 65–75.

[68] Gideon Meyerowitz-Katz, Sumathy Ravi, Leonard Arnolda, Xiaoqi Feng, Glen Maberly, and Thomas Astell-Burt. 2020. Rates of attrition and dropout in app-based interventions for chronic disease: systematic review and meta-analysis. *Journal of Medical Internet Research* 22, 9 (2020), e20283.

An Analysis of User Reviews of Mobile Mental Health Applications                                                                                     421:27[69] Adam S Miner, Arnold Milstein, Stephen Schueller, Roshini Hegde, Christina Mangurian, and Eleni Linos. 2016. Smartphone-based conversational agents and responses to questions about mental health, interpersonal violence, and physical health. *JAMA internal medicine* 176, 5 (2016), 619–625.

[70] Tally Moses. 2009. Self-labeling and its effects among adolescents diagnosed with mental disorders. *Social science & medicine* 68, 3 (2009), 570–578.

[71] Susan M Mudambi and David Schuff. 2010. Research note: What makes a helpful online review? A study of customer reviews on Amazon. com. *MIS quarterly* (2010), 185–200.

[72] Cosmin Munteanu, Heather Molyneaux, and Susan O'Donnell. 2014. Fieldwork with vulnerable populations. *Interactions* 21, 1 (2014), 50–53.

[73] P Muppirishetty and Minha Lee. 2020. Voice user interfaces for mental healthcare: leveraging technology to help our inner voice. In *3rd ACM Conference on Computer-Supported Cooperative Work and Social Computing, CSCW 2020*.

[74] Michelle M Ng, Joseph Firth, Mia Minen, and John Torous. 2019. User engagement in mental health apps: a review of measurement, reporting, and validity. *Psychiatric Services* 70, 7 (2019), 538–544.

[75] J Nicholas, K Boydell, and H Christensen. 2017. Beyond symptom monitoring: Consumer needs for bipolar disorder self-management using smartphones. *European Psychiatry* 44 (2017), 210–216.

[76] Jennifer Nicholas, Andrea S Fogarty, Katherine Boydell, and Helen Christensen. 2017. The reviews are in: a qualitative content analysis of consumer perspectives on apps for bipolar disorder. *Journal of medical Internet research* 19, 4 (2017), e105.

[77] Lorelli S Nowell, Jill M Norris, Deborah E White, and Nancy J Moules. 2017. Thematic analysis: Striving to meet the trustworthiness criteria. *International journal of qualitative methods* 16, 1 (2017), 1609406917733847.

[78] Kyo-Joong Oh, Dongkun Lee, Byungsoo Ko, and Ho-Jin Choi. 2017. A chatbot for psychiatric counseling in mental healthcare service based on emotional dialogue analysis and sentence generation. In *2017 18th IEEE International Conference on Mobile Data Management (MDM)*. IEEE, 371–375.

[79] Kathleen O'Leary, Arpita Bhattacharya, Sean A. Munson, Jacob O. Wobbrock, and Wanda Pratt. 2017. Design Opportunities for Mental Health Peer Support Technologies. In *Proceedings of the 2017 ACM Conference on Computer Supported Cooperative Work and Social Computing* (Portland, Oregon, USA) *(CSCW '17)*. Association for Computing Machinery, New York, NY, USA, 1470–1484. https://doi.org/10.1145/2998181.2998349

[80] Kathleen O'Leary, Stephen M Schueller, Jacob O Wobbrock, and Wanda Pratt. 2018. "Suddenly, we got to become therapists for each other" Designing Peer Support Chats for Mental Health. In *Proceedings of the 2018 CHI Conference on Human Factors in Computing Systems*. 1–14.

[81] title =The Mental Health App Guide Designed With You In Mind url =https://onemindpsyberguide.org/about-psyberguide/ lastaccessed =August 25, 2021 One Mind PsyberGuide, year = 2021. [n.d.]. .

[82] World Health Organization. 2001. The World Health Report 2001: Mental health: new understanding, new hope. (2001).

[83] Dennis Pagano and Walid Maalej. 2013. User feedback in the appstore: An empirical study. In *2013 21st IEEE international requirements engineering conference (RE)*. IEEE, 125–134.

[84] Sepideh Paknejad. 2018. Sentiment classification on Amazon reviews using machine learning approaches.

[85] Leysia Palen and Stinne Aaløkke. 2006. Of pill boxes and piano benches: " home-made" methods for managing medication. In *Proceedings of the 2006 20th anniversary conference on Computer supported cooperative work*. 79–88.

[86] Lisa Parker, Tanya Karliychuk, Donna Gillies, Barbara Mintzes, Melissa Raven, and Quinn Grundy. 2017. A health app developer's guide to law and policy: a multi-sector policy analysis. *BMC medical informatics and decision making* 17, 1 (2017), 1–13.

[87] Marie-Camille Patoz, Diego Hidalgo-Mazzei, Bruno Pereira, Olivier Blanc, Ingrid de Chazeron, Andrea Murru, Norma Verdolini, Isabella Pacchiarotti, Eduard Vieta, Pierre-Michel Llorca, et al. 2021. Patients' adherence to smartphone apps in the management of bipolar disorder: a systematic review. *International journal of bipolar disorders* 9, 1 (2021), 1–15.

[88] Sachin R. Pendse, Kate Niederhoffer, and Amit Sharma. 2019. Cross-Cultural Differences in the Use of Online Mental Health Support Forums. *Proc. ACM Hum.-Comput. Interact.* 3, CSCW, Article 67 (nov 2019), 29 pages. https://doi.org/10.1145/3359169

[89] Sara Pinto, Sílvia Caldeira, and José Carlos Martins. 2017. e-Health in palliative care: review of literature, Google Play and App Store. *International journal of palliative nursing* 23, 8 (2017), 394–401.

[90] Elisabeth Platzer. 2011. Opportunities of automated motive-based user review analysis in the context of mobile app acceptance. In *Central European Conference on Information and Intelligent Systems*. Faculty of Organization and Informatics Varazdin, 309.

[91] Mary Ann Priester, Teri Browne, Aidyn Iachini, Stephanie Clone, Dana DeHart, and Kristen D Seay. 2016. Treatment access barriers and disparities among individuals with co-occurring mental health and substance use disorders: an integrative literature review. *Journal of substance abuse treatment* 61 (2016), 47–59.
Proc. ACM Hum.-Comput. Interact., Vol. 6, No. CSCW2, Article 421. Publication date: November 2022.

[92] Sulakshan Rasiah and Jonathan K Kam. 2013. Clinical software on personal mobile devices needs regulation. *The Medical Journal of Australia* 198, 10 (2013), 530–531.
[93] Sabirat Rubya. 2017. Facilitating Peer Support for Recovery from Substance Use Disorders *(CHI EA '17)*. Association for Computing Machinery, New York, NY, USA. https://doi.org/10.1145/3027063.3048431
[94] Sabirat Rubya, Joseph Numainville, and Svetlana Yarosh. 2021. Comparing Generic and Community-Situated Crowd-sourcing for Data Validation in the Context of Recovery from Substance Use Disorders. In *Proceedings of the 2021 CHI Conference on Human Factors in Computing Systems* (Yokohama, Japan) *(CHI '21)*. Association for Computing Machinery, New York, NY, USA, Article 449, 17 pages. https://doi.org/10.1145/3411764.3445399
[95] Sabirat Rubya, Xizi Wang, and Svetlana Yarosh. 2019. HAIR: Towards Developing a Global Self-Updating Peer Support Group Meeting List Using Human-Aided Information Retrieval. In *Proceedings of the 2019 Conference on Human Information Interaction and Retrieval* (Glasgow, Scotland UK) *(CHIIR '19)*. Association for Computing Machinery, New York, NY, USA, 83–92. https://doi.org/10.1145/3295750.3298933
[96] Sabirat Rubya and Svetlana Yarosh. 2017. Interpretations of online anonymity in Alcoholics Anonymous and Narcotics Anonymous. *Proceedings of the ACM on Human-Computer Interaction* 1, CSCW (2017), 1–22.
[97] Sabirat Rubya and Svetlana Yarosh. 2017. Video-mediated peer support in an online community for recovery from substance use disorders. In *Proceedings of the 2017 ACM Conference on Computer Supported Cooperative Work and Social Computing*. 1454–1469.
[98] Jasmina Rueger, Wilfred Dolfsma, and Rick Aalbers. 2021. Perception of peer advice in online health communities: Access to lay expertise. *Social Science & Medicine* 277 (2021), 113117.
[99] Koustuv Saha, Sindhu Kiranmai Ernala, Sarmistha Dutta, Eva Sharma, and Munmun De Choudhury. 2020. Understanding moderation in online mental health communities. In *International Conference on Human-Computer Interaction.* Springer, 87–107.
[100] Danielle Schlosser, Timothy Campellone, Daniel Kim, Brandy Truong, Silvia Vergani, Charlie Ward, and Sophia Vinogradov. 2016. Feasibility of PRIME: a cognitive neuroscience-informed mobile app intervention to enhance motivated behavior and improve quality of life in recent onset schizophrenia. *JMIR research protocols* 5, 2 (2016), e5450.
[101] Nelson Shen, Michael-Jane Levitan, Andrew Johnson, Jacqueline Lorene Bender, Michelle Hamilton-Page, Alejandro Alex R Jadad, and David Wiljer. 2015. Finding a depression app: a review and content analysis of the depression app marketplace. *JMIR mHealth and uHealth* 3, 1 (2015), e3713.
[102] Tanner Skousen, Hani Safadi, Colleen Young, Elena Karahanna, Sami Safadi, and Fouad Chebib. 2020. Successful moderation in online patient communities: inductive case study. *Journal of medical Internet research* 22, 3 (2020), e15983.
[103] Petr Slovák, Nikki Theofanopoulou, Alessia Cecchet, Peter Cottrell, Ferran Altarriba Bertran, Ella Dagan, Julian Childs, and Katherine Isbister. 2018. " I just let him cry... Designing Socio-Technical Interventions in Families to Prevent Mental Health Disorders. *Proceedings of the ACM on Human-Computer Interaction* 2, CSCW (2018), 1–34.
[104] Dayana Spagnuelo, Ana Ferreira, and Gabriele Lenzini. 2019. Accomplishing Transparency within the General Data Protection Regulation.. In *ICISSP*. 114–125.
[105] Kamonphop Srisopha, Daniel Link, Devendra Swami, and Barry Boehm. 2020. Learning Features that Predict Developer Responses for iOS App Store Reviews. In *Proceedings of the 14th ACM/IEEE International Symposium on Empirical Software Engineering and Measurement (ESEM)*. 1–11.
[106] Timo Stolz, Ava Schulz, Tobias Krieger, Alessia Vincent, Antoine Urech, Christian Moser, Stefan Westermann, and Thomas Berger. 2018. A mobile app for social anxiety disorder: A three-arm randomized controlled trial comparing mobile and PC-based guided self-help interventions. *Journal of Consulting and Clinical Psychology* 86, 6 (2018), 493.
[107] Elizabeth Stowell, Mercedes C Lyson, Herman Saksono, Reneé C Wurth, Holly Jimison, Misha Pavel, and Andrea G Parker. 2018. Designing and evaluating mHealth interventions for vulnerable populations: A systematic review. In *Proceedings of the 2018 CHI Conference on Human Factors in Computing Systems*. 1–17.
[108] Ali Sunyaev, Tobias Dehling, Patrick L Taylor, and Kenneth D Mandl. 2015. Availability and quality of mobile health app privacy policies. *Journal of the American Medical Informatics Association* 22, e1 (2015), e28–e33.
[109] Felix Ter Chian Tan and Rajesh Vasa. 2011. Toward a social media usage policy. In *Proc. ACIS*. 84–89.
[110] Kong Saoane Thach. 2019. A Qualitative Analysis of User Reviews on Mental Health Apps: Who Used it? for What? and Why?. In *2019 IEEE-RIVF International Conference on Computing and Communication Technologies (RIVF)*. IEEE, 1–4.
[111] Val Theisz. 2019. *Medical device regulatory practices: an international perspective.* Pan Stanford.
[112] John Torous, Jennifer Nicholas, Mark E Larsen, Joseph Firth, and Helen Christensen. 2018. Clinical review of user engagement with mental health smartphone apps: evidence, theory and improvements. *Evidence-based mental health* 21, 3 (2018), 116–119.

[113] John Torous and Laura Weiss Roberts. 2017. Needed innovation in digital health and smartphone applications for mental health: transparency and trust. *JAMA psychiatry* 74, 5 (2017), 437–438.
[114] John Blake Torous, Steven Richard Chan, Shih Yee-Marie Tan Gipson, Jung Won Kim, Thuc-Quyen Nguyen, John Luo, and Philip Wang. 2018. A hierarchical framework for evaluation and informed decision making regarding smartphone apps for clinical care. *Psychiatric Services* 69, 5 (2018), 498–500.
[115] Aditya Nrusimha Vaidyam, Hannah Wisniewski, John David Halamka, Matcheri S Kashavan, and John Blake Torous. 2019. Chatbots and conversational agents in mental health: a review of the psychiatric landscape. *The Canadian Journal of Psychiatry* 64, 7 (2019), 456–464.
[116] Rajesh Vasa, Leonard Hoon, Kon Mouzakis, and Akihiro Noguchi. 2012. A preliminary analysis of mobile app user reviews. In *Proceedings of the 24th Australian computer-human interaction conference*. 241–244.
[117] John Vines, Roisin McNaney, Rachel Clarke, Stephen Lindsay, John McCarthy, Steve Howard, Mario Romero, and Jayne Wallace. 2013. Designing for-and with-vulnerable people. In *CHI'13 Extended Abstracts on Human Factors in Computing Systems*. 3231–3234.
[118] Jennifer K Wagner. 2020. The Federal Trade Commission and Consumer Protections for Mobile Health Apps. *Journal of Law, Medicine & Ethics* 48, S1 (2020), 103–114.
[119] Xiaomei Wang, Carl Markert, and Farzan Sasangohar. 2021. Investigating Popular Mental Health Mobile Application Downloads and Activity During the COVID-19 Pandemic. *Human Factors* (2021), 0018720821998110.
[120] Carolyn Windler, Maeve Clair, Cassandra Long, Leah Boyle, and Ana Radovic. 2019. Role of moderators on engagement of adolescents with depression or anxiety in a social media intervention: content analysis of web-based interactions. *JMIR mental health* 6, 9 (2019), e13467.
[121] Qiang Ye, Ziqiong Zhang, and Rob Law. 2009. Sentiment classification of online reviews to travel destinations by supervised machine learning approaches. *Expert systems with applications* 36, 3 (2009), 6527–6535.
[122] Michael Zimmer. 2010. "But the data is already public": on the ethics of research in Facebook. *Ethics and information technology* 12, 4 (2010), 313–325.